\documentclass[11pt,english]{extarticle}

\usepackage[T1]{fontenc}
\usepackage[latin9]{inputenc}
\usepackage[skip=\medskipamount]{parskip}
\usepackage{color}
\usepackage{amsmath}
\usepackage{amssymb}
\usepackage{graphicx}
\usepackage{geometry}
\makeatletter
\@ifundefined{date}{}{\date{}}

\newtheorem{remark}{Remark}

\newcommand{\vs}{\vspace}
\newcommand{\sis}{\setlength{\itemsep}}

\date{}

\makeatother

\usepackage{babel}
\begin{document}
\title{A two-stage approach to satellite constellation optimization:\\classical
and QUBO formulations\thanks{The research presented in this paper has been carried out under the
project Space It Up! - Spoke 1: enabling technologies for novel near-Earth
and exploration missions, funded by the Italian Space Agency (ASI).}}
\author{Carlo Novara\\Politecnico di Torino (Italy)\\Department of Electronics
and Telecommunications\\{\small Email: }{\small\texttt{carlo.novara@polito.it}}}
\maketitle
\begin{abstract}
The design of satellite constellations for Earth observation requires
balancing spatial coverage, revisit time, cost, and operational complexity.
This paper considers the problem of designing the orbits of a given
number of Low Earth Orbit (LEO) or Very Low Earth Orbit (VLEO) satellites
to maximize the spatial and temporal resolution achieved over a prescribed
set of ground targets. This kind of problem is inherently nonconvex
and possibly combinatorial, making its solution computationally demanding
for large constellations and target sets. To address this challenge,
we propose a two-stage optimization strategy that separates spatial-coverage
design from temporal-resolution optimization, thereby reducing the
complexity of the overall problem. Two variants of the method are
developed. The first employs continuous decision variables during
the spatial-optimization stage, whereas the second discretizes these
variables and reformulates the problem as a Quadratic Unconstrained
Binary Optimization (QUBO) problem. This latter enables the use of
efficient QUBO solvers and leads to the formulation an optimization
problem whose solution can be written in closed-form. The proposed
framework provides a scalable approach to the design of heterogeneous
LEO and VLEO Earth-observation constellations and establishes a pathway
to new methods in satellite mission design. Preliminary simulation
results are presented to demonstrate the effectiveness of the strategy.
\end{abstract}

\section{Introduction}

Constellation optimization is essential for managing emerging systems
composed of Low Earth Orbit (LEO) and Very Low Earth Orbit (VLEO)
satellites dedicated to Earth Observation. As demand grows for imagery
and data products with both very high spatial resolution and very
high temporal resolution, individual satellites or single homogeneous
constellations are no longer sufficient. Instead, future Earth Observation
architectures will increasingly rely on coordinated networks of satellites
operating across different orbits, carrying different sensors, and
serving multiple users with competing priorities.

In this context, optimization is needed to determine how satellites
should be designed, deployed, operated, and coordinated. For LEO and
VLEO missions, key decisions include orbital altitude, inclination,
phasing, revisit geometry, ground-track repeatability, sensor allocation,
downlink opportunities, propulsion usage, and collision-risk mitigation.
These decisions are strongly coupled: for example, increasing the
number of satellites can improve revisit time, but it also increases
operational complexity, cost, communication demand, and space-traffic-management
constraints.

The type of problems considered in this paper is as follows. A set
of $N_{T}$ targets located on a given region of the Earth surface
must be monitored by a constellation of $N_{S}$ satellites. The problem
we consider is the following: Given a fixed number of satellites $N_{S}$,
design the constellation orbits ensuring a monitoring of the $N_{T}$
targets with the highest possible spatial and temporal resolutions.
A strictly related problem is: Design a constellation with a minimum
number $N_{S}$ of satellites, ensuring a monitoring of the targets
of interest with prescribed spatial and temporal resolutions. 

The solution of such problems requires in general an exponential or
combinatorial computational complexity, and using standard algorithms
running on standard computers may require extremely long computation
times or not give satisfactory results.

Several works are available in the literature about orbit design for
large constellations/formations, see, e.g., \cite{const_ps_2024,Crisp2019,Deccia2022,Imoto2023,Kim2019,Melaku2023,Nunes2013,Thangavel2024}.
See also the surveys \cite{Curzi2020,Zhang2022} and the references
therein. In these works, the considered orbit optimization and guidance
problems are tackled using well-assessed nonlinear optimization techniques,
including genetic algorithms, particle swarm, ant colony optimization,
heuristic search, with a prevalence of genetic algorithms. Several
papers can also be found about constellation/formation reconfiguration
and space traffic management \cite{Curzi2020,Appel2014,Lagona2022,Stoilova2022,Arnas2022,Chen2024,Fakoor2016,Jiaxin2021,Soleymani2019}.
Also in these latter works, the involved guidance/trajectory planning
problems are tackled by means of nonlinear optimization techniques,
including genetic algorithms, particle swarm, nonlinear direct and
indirect methods.

In the present paper, we propose a novel optimization approach based
on a two-stage strategy, that may allow a significant reduction of
the computation time needed to design optimal constellations. In the
first stage, we optimize the spatial coverage, while in the second
stage we optimize the temporal resolution. Two variants of the approach
are developed. In the first variant, the decision variables of the
first stage are continuous variables. In the second variant, the decision
variables of the first stage are discretized, allowing us to obtain
a Quadratic Unconstrained Binary Optimization (QUBO) formulation of
the design problem. This formulation enables the use of efficient
QUBO solvers and allows a significant simplification of the optimization
problem, yielding a new formulation for which closed-form solutions
can be found.  Preliminary numerical experiments are carried out
to demonstrate the effectiveness of the proposed strategy.

\section{Preliminaries }

\subsection{Acronyms}

AP: Argument of Perigee

ECI: Earth-Centered Inertial 

LEO: Low Earth Orbit

LP: Linear Programming

PF: PeriFocal

RAAN: Right Ascension of the Ascending Node

RGT: Repeat Ground-Track

TA: True Anomaly

VLEO: Very Low Earth Orbit

w.r.t.: with respect to

\subsection{Main parameters/variables}

$N_{S}$: total number of satellites

$n_{S}$: number of satellites per orbit

$N_{O}$: number of orbits

$N_{T}$: number of targets

$N_{R}$: number of orbit revolutions

$P_{O}$: orbit period

$T_{\mathrm{sid}}$: sidereal day

$\omega_{E}$: Earth rotation speed

$a$: semi-major axis

$\mathfrak{i}$: inclination (gothic style used to distinguish it
from index $i$)

$\vec{e}$: eccentricity vector

$e$: eccentricity

$\Omega$: right ascension of the ascending node

$\omega$: argument of perigee

$\lambda$: longitude

$\phi$: latitude

$J2000$ epoch: January 1, 2000, at 12:00 Terrestrial Time (TT)

\subsection{Notation}
\begin{itemize}
\item \sis{1mm}Scalars: $a,b,A,B\in\mathbb{R}$.
\item Column vectors: \vs{-4mm}
\[
r=\left[\begin{array}{c}
r_{1}\\
\vdots\\
r_{n}
\end{array}\right]=[r_{1}\;\ldots\;r_{n}]^{\top}=(r_{1},\ldots,r_{n})\in\mathbb{R}^{n\times1}.
\]
\vs{-4mm}
\item Row vectors: $r^{\top}=[r_{1}\;\ldots\;r_{n}]\in\mathbb{R}^{1\times n}$.
\item Matrices: $M=\left[\begin{array}{ccc}
M_{11} & \cdots & M_{1m}\\
\vdots & \ddots & \vdots\\
M_{n1} & \cdots & M_{nm}
\end{array}\right]=[M_{ij}]\in\mathbb{R}^{n\times m}$. 
\begin{itemize}
\item $j$th row of a matrix: $M_{j:}$; $i$th column of a matrix: $M_{:i}$.
\item $\mathbf{1}_{n\times m}$ is a matrix of dimensions $n\times m$ with
all elements equal to $1$. $\mathbf{1}_{n}$ is a square matrix of
dimensions $n\times n$ with all elements equal to $1$. $I_{n}$
is the identity matrix of dimension $n$. When not necessary, the
subscript $n$ is omitted.
\end{itemize}
\item Sets: $\{a_{1,},a_{2},\ldots,a_{n}\}=\{a_{i}\}$, $\{a_{11,},a_{12},\ldots,a_{21},a_{22},\ldots,a_{nm}\}=\{a_{ij}\}$.
\item Vector $\ell_{2}$ (Euclidean) norm: $\left\Vert r\right\Vert \equiv\left\Vert r\right\Vert _{2}\doteq\begin{array}{c}
\sqrt{\sum_{i=1}^{n}r_{i}^{2}}.\end{array}$
\item Geodesic distance: Consider two points located on the surface of a
sphere with radius $R$. The longitude and latitude of the two points
are given by $(\lambda,\phi)$ and $(\bar{\lambda},\bar{\phi})$,
respectively. Define $\mathfrak{u}(\lambda,\phi)\doteq(\cos\phi\cos\lambda,\cos\phi\sin\lambda,\sin\phi)\in\mathbb{R}^{3\times1}$.
Their geodesic distance is defined as
\[
\left|(\lambda,\phi)-(\bar{\lambda},\bar{\phi})\right|_{R}\doteq R\,\arccos(\mathfrak{u}(\lambda,\phi)^{\top}\mathfrak{u}(\bar{\lambda},\bar{\phi})).
\]
 
\end{itemize}

\subsection{Reference frames}

The following reference frames are considered.
\begin{itemize}
\item \sis{0mm}Perifocal (PF) frame: origin at the Earth CoM; axes defined
by the following unit vectors:\vs{0.7mm}
\begin{itemize}
\item \sis{0.7mm}$\mathbf{p}_{1}=\vec{e}/e$: normalized eccentricity vector
passing through the periapsis, on the orbit plane;
\item $\mathbf{p}_{3}$: orbit pole (perpendicular to the orbit plane);
\item $\mathbf{p}_{2}=\mathbf{p}_{3}\times\mathbf{p}_{1}$: on the orbital
plane. 
\end{itemize}
\item Earth-Centered Inertial (ECI) frame J2000: origin at the Earth CoM;
axes defined by the following unit vectors:\vs{0.7mm}
\begin{itemize}
\item \sis{0.7mm}$\hat{\mathbf{I}}$: J2000 mean vernal equinox direction;
\item $\hat{\mathbf{K}}$: polar rotation axis;
\item $\hat{\mathbf{J}}=\hat{\mathbf{K}}\times\hat{\mathbf{I}}$: on the
equatorial plane. 
\end{itemize}
\end{itemize}

\section{Problem setting}

\subsection{Orbit description}

An orbit is univocally determined by $6$ orbital elements: semi-major
axis $a$, eccentricity $e$, inclination $\mathfrak{i}$, Right Ascension
of the Ascending Node (RAAN) $\Omega$, Argument of Perigee (AP) $\omega$,
true anomaly (TA) $\theta$.

In our setting, the orbit of a satellite is described by the equation
\begin{equation}
r=\frac{p}{1+e\cos\theta}\label{eq:ore}
\end{equation}
where $r$ is the satellite distance from the Earth center of mass
and $p=a(1-e^{2})$ is the semilatus rectum of the orbit. Thanks to
its simplicity, the orbit equation (\ref{eq:ore}) is well suited
for optimization purposes. The orbits obtained by means of optimization
processes based on (\ref{eq:ore}) will be validated using a high-fidelity
simulator, describing phenomena like Earth gravity harmonics, atmospheric
drag, solar radiation pressure, and other perturbations.

Equation (\ref{eq:ore}) provides a geometric description of the orbit
in the Perifocal frame. In particular, the satellite position in this
frame is 
\[
\begin{array}{l}
\mathbf{r}_{P}=\left[\begin{array}{c}
x_{P}\\
y_{P}\\
z_{P}
\end{array}\right]=\left[\begin{array}{c}
r\cos\theta\\
r\sin\theta\\
0
\end{array}\right].\end{array}
\]
The satellite position in the ECI frame is
\[
\begin{array}{l}
\mathbf{r}=\left[\begin{array}{c}
x\\
y\\
z
\end{array}\right]=\mathbf{T}_{313}(\Omega,\mathfrak{i},\omega)\mathbf{r}_{P}\end{array}
\]
where $\mathbf{T}_{313}(\Omega,\mathfrak{i},\omega)$ is a 313 rotation
matrix with angles $\Omega,\mathfrak{i},\omega$.  

\subsection{Ground-track}

The ground-track of an orbit is the vertical projection onto the surface
of the rotating Earth of the path traced by a satellite traveling
on the orbit. Longitude and latitude of this projection are
\begin{equation}
\begin{alignedat}{1}\lambda & =\mathrm{atan2}(y,x)-\omega_{E}\,t\\
\phi & =\mathrm{asin}(z,r)
\end{alignedat}
\label{eq:gt_lonlat}
\end{equation}
where $\omega_{E}$ is the Earth angular speed and $t$ is the time.
 The ground-track is the curve on the Earth surface defined as
\[
\Gamma\doteq\{(\lambda(t),\phi(t)):t\in[0,T_{f}]\}
\]
where $[0,T_{f}]$ is a time interval of interest.

\subsection{Coverage }

A set of $N_{T}$ fixed targets located on a given region of the Earth
surface and a constellation of $N_{S}$ satellites are given. For
simplicity, the target with index $j$ is called target $j$, while
the satellite with index $l$ is called satellite $l$. For every
satellite $j$, we assume that the initial value of the true anomaly
in (\ref{eq:ore}) is $\theta_{j}(0)=0$.

The longitude and latitude of target $j$ are denoted by $\bar{\lambda}_{j}$
and $\bar{\phi}_{j}$, respectively. The longitude and latitude of
the ground-track of satellite $l$ at a given time $t$ are denoted
by $\lambda_{l}(t)$ and $\phi_{l}(t)$, respectively.

We say that target $j$ is visited (or observed) by satellite $l$
in the time interval $[0,T_{f}]$ if 
\[
d_{jl}(t)\doteq\left|(\lambda_{l}(t),\phi_{l}(t))-(\bar{\lambda}_{j},\bar{\phi}_{j})\right|_{R_{E}}\leq s,\quad\mathrm{for\;some}\;t\in[0,T_{f}]
\]
where $\left|\cdot\right|_{R_{E}}$ is the geodesic distance and $s$
is a threshold that depends on the observation instruments mounted
on the satellite. 

In a given time interval $[0,T_{f}]$, a target may be visited more
than once. The number of times in which a target $j$ is visited by
a satellite $l$ in a time interval $[0,T_{f}]$ is
\begin{equation}
M_{jl}=\mathrm{card}(\mathcal{T}_{jl})\label{eq:visit_count}
\end{equation}
where $\mathrm{card}(\cdot)$ is the set cardinality and
\[
\begin{alignedat}{1}\mathcal{T}_{jl} & \doteq\{\tau\in[0,T_{f}]:d_{jl}(\tau)=s,\dot{d}_{jl}(\tau)<0\}\end{alignedat}
\]
is the set of time instants at which a satellite $l$ starts a single
observation of a target $j$. The matrix $M\doteq[M_{jl}]$ is called
Visit-Count (VC) matrix.

The count of all visits that a target $j$ receives by all satellites
of the constellation in the interval $[0,T_{f}]$ is given by
\begin{equation}
\mathcal{N}_{j}\doteq\sum_{l=1}^{N_{s}}M_{jl}.\label{eq:Nj}
\end{equation}
The mean and minimum coverages of the constellation in the interval
$[0,T_{f}]$ are respectively defined as
\begin{equation}
\begin{alignedat}{1} & C_{\mathrm{mean}}\doteq\frac{1}{N_{T}}\sum_{j=1}^{N_{T}}\mathcal{N}_{j}\\
 & C_{\mathrm{min}}\doteq\min_{j=1,\ldots,N_{T}}\mathcal{N}_{j}.
\end{alignedat}
\label{eq:cove1}
\end{equation}

\subsection{Revisit time}

In our setting, we consider circular RGT orbits, i.e., orbits whose
ground-track on the rotating Earth repeats exactly after a fixed number
of days, meaning the satellite passes over the same points on the
Earth\textquoteright s surface at regular intervals. RGT orbits are
widely used for Earth observation since they allow systematic acquisitions
and make it easy to monitor changes, e.g., vegetation growth, urban
expansion, glacier retreat or disaster assessment.  Therefore, the
time interval of interest for our constellation design is $[0,T_{f}]$,
where 
\[
T_{f}=N_{R}P_{O}=N_{\mathrm{sid}}T_{\mathrm{sid}},
\]
$N_{R}$ is the number of revolutions in an integer number $N_{\mathrm{sid}}$
of sidereal days, $P_{O}$ is the orbit period and $T_{\mathrm{sid}}$
is the sidereal day ($\sim86164\;\mathrm{s}$). In the following,
we assume the time interval $[0,T_{f}]$ and not indicate it explicitly
if not necessary. 

Assuming $\mathcal{N}_{j}>0,\;\forall j$, we define the following
sets of time instants:
\begin{equation}
\begin{alignedat}{1} & \breve{\mathcal{T}}_{j}\doteq\mathrm{sort}\left\{ \bigcup_{l=1}^{N_{S}}\mathcal{T}_{jl}\right\} \\
 & \bar{\mathcal{T}}_{j}\doteq\breve{\mathcal{T}}_{j}\cup\{T_{\mathrm{sid}}+\breve{\mathcal{T}}_{j1}\}
\end{alignedat}
\label{eq:Tj}
\end{equation}
where $\mathrm{sort}(\cdot)$ is the ascending-order sorting operator
and $\breve{\mathcal{T}}_{j1}$ is the first element of $\breve{\mathcal{T}}_{j}$.
Note that the set $\breve{\mathcal{T}}_{j}$ contains all the time
instants in which a target $j$ is visited (organized in ascending
order). The set $\bar{\mathcal{T}}_{j}$ is obtained from $\breve{\mathcal{T}}_{j}$,
including the first visiting instant of the subsequent day. According
to (\ref{eq:Nj}) and (\ref{eq:Tj}), the set $\bar{\mathcal{T}}_{j}$
contains a finite number of time instants, i.e., 
\begin{equation}
\bar{\mathcal{T}}_{j}=\{t_{j,1},t_{j,2},\ldots,t_{j,\mathcal{N}_{j}+1}\}.\label{eq:Tb}
\end{equation}
The set of revisit times for target $j$ is thus given by
\[
\Delta_{j}\doteq\{\delta{}_{j,1},\delta{}_{j,2},\ldots,\delta_{j,\mathcal{N}_{j}}\}
\]
where $\delta{}_{j,i}=t_{j,i+1}-t_{j,i}$, $i=1,2,\ldots,\mathcal{N}_{j}$.
If $\mathcal{N}_{j}=0$ for some $j$, we set $\Delta_{j}\doteq\{\infty\}$.

The mean and maximum revisit times of the constellation in the interval
$[0,T_{f}]$ are respectively defined as
\begin{equation}
\begin{alignedat}{1} & \Delta_{\mathrm{mean}}\doteq\frac{1}{N_{T}}\sum_{j=1}^{N_{T}}\frac{1}{\mathcal{N}_{j}}\sum_{i=1}^{\mathcal{N}_{j}}\delta{}_{j,i}\\
 & \Delta_{\mathrm{max}}\doteq\max_{j=1:N_{T}}\,\max_{i=1:\mathcal{N}_{j}}\,\delta{}_{j,i}.
\end{alignedat}
\label{eq:rvt1}
\end{equation}

\subsection{Problem formulation}

A set of $N_{T}$ targets located on a given region of the Earth surface
must be monitored by a constellation of $N_{S}$ satellites traveling
on prescribed orbits. As discussed above, we focus on circular RGT
orbits, i.e., orbits whose ground-track on the rotating Earth repeats
exactly after a fixed number of days. 

We recall that the orbit of each satellite is univocally determined
by $6$ orbital elements: semi-major axis $a$, eccentricity $e$,
inclination $\mathfrak{i}_{l}$, Right Ascension of the Ascending
Node (RAAN) $\Omega_{l}$, Argument of Perigee (AP) $\omega_{l}$,
true anomaly (TA) $\theta_{l}$, where $l$ is the index indicating
the satellite. In circular orbits, $\omega_{l}$ are $\theta_{l}$
are individually undefined but their sum is well-defined. In our setting,
to avoid ill-defined orbits, we assume null APs ($\omega_{l}=0,\;\forall l$)
allowing us to have well-defined TAs. In the case of equatorial circular
orbit, we assume $\omega_{l}=0$, $\Omega_{l}=0$, $\forall l$.

Therefore, the orbital elements of the considered orbits are as follows:
\begin{itemize}
\item Semi-major axis (fixed and equal for all satellites): $a\in[R_{E}+600,R_{E}+800]$,
where $R_{E}=6378\;\mathrm{km}$ is the average Earth radius. Note
that $a$ depends on $N_{R}$ and $N_{\mathrm{sid}}$.
\item Eccentricity (fixed and equal for all satellites): $e=0$.
\item Inclinations $\mathfrak{i}_{l}\in D_{\mathfrak{i}}\doteq[0,\pi]$,
$l=1,\ldots,N_{S}$: decision variables.
\item RAANs $\Omega_{l}\in D_{\Omega}\doteq[0,2\pi]$, $l=1,\ldots,N_{S}$:
decision variables.
\item Initial TAs $\theta_{0l}\doteq\theta_{l}(0)\in\mathcal{D}_{\theta}\doteq[0,2\pi]$,
$l=1,\ldots,N_{S}$: decision variables.
\end{itemize}
The core problem we want to solve is: Given $N_{T}$ targets and $N_{S}$
satellites, design the constellation orbits ensuring a monitoring
of the targets with the highest possible spatial and temporal resolutions.
To this aim, we define the multi-objective function
\begin{equation}
J_{0}(\mathbf{i},\mathbf{\Omega},\boldsymbol{\theta}_{0})\doteq-\mu_{1}C_{\mathrm{mean}}-\mu_{2}C_{\mathrm{min}}+\mu_{3}\Delta_{\mathrm{mean}}+\mu_{4}\Delta_{\mathrm{max}}\label{eq:J0}
\end{equation}
where $\mathbf{i}=(\mathfrak{i}_{1},\ldots,\mathfrak{i}_{N_{S}})$,
$\mathbf{\Omega}=(\Omega_{1},\ldots,\Omega_{N_{S}})$ and $\boldsymbol{\theta}_{0}=(\theta_{01},\ldots,\theta_{0N_{S}})$
are the the decision variables, and $\mu_{i}$ are weights chosen
depending on the goals of the constellation design. An optimal constellation
design is obtained by solving the optimization problem
\begin{equation}
\begin{alignedat}{1} & (\mathbf{i}^{*},\mathbf{\Omega}^{*},\boldsymbol{\theta}_{0}^{*})=\arg\min_{\mathbf{i},\mathbf{\Omega},\boldsymbol{\theta}_{0}}J_{0}(\mathbf{i},\mathbf{\Omega},\boldsymbol{\theta}_{0})\\
 & \mathrm{subject\;to}:\\
 & \mathbf{i}\in D_{\mathfrak{i}}^{N_{S}},\;\mathbf{\Omega}\in D_{\Omega}^{N_{S}},\;\boldsymbol{\theta}_{0}\in\mathcal{D}_{\theta}^{N_{S}}.
\end{alignedat}
\label{eq:opt0}
\end{equation}
Minimizing $\Delta_{\alpha}$, with $\alpha\in\{\mathrm{mean},\mathrm{max}\}$,
allows us to improve the constellation temporal resolution and, indirectly,
to maximize the coverage.\textcolor{red}{{} }Solving problem (\ref{eq:opt0})
for different $N_{S}$, we can design a constellation with a minimum
number $N_{S}$ of satellites, ensuring a monitoring of the $N_{T}$
targets with a prescribed temporal resolution. Creating a grid with
$N_{T}$ targets and increasing $N_{T}$ allows us to improve the
spatial resolution.

\begin{remark}\label{rem:comp_compl_0} Problem (\ref{eq:opt0})
is nonlinear and non-convex. Finding a global solution involves an
exponential computational complexity. Therefore, even finding a local
solution may require extremely long computation times or not give
satisfactory results. Local solutions of problem (\ref{eq:opt0})
can be found using standard optimization algorithms like particle
swarm, genetic algorithms, simulated annealing, surrogate optimization.
$\square$ \end{remark}

\section{Two-stage constellation optimization}

\label{sec:2st_optim}

In this section, we propose a two-stage approach for constellation
optimization, allowing us to reduce the computational complexity w.r.t.
standard nonlinear algorithms used to solve problems like (\ref{eq:opt0}).
In the first stage, the coverage of the constellation is maximized
w.r.t. to the orbit inclinations and RAANs. In the second stage, given
the optimal orbits obtained in the first stage, the revisit times
are minimized w.r.t. the initial TAs. 

\subsection{Stage 1: Coverage maximization}

In stage 1, the goal is to maximize the coverage of the constellation
ground-track, which depends on the number of orbits $N_{O}$, and
on their inclinations and RAANs. Hence, we define the objective function
\begin{equation}
J_{C}(\mathbf{i},\mathbf{\Omega})\doteq-\mu_{1}C_{\mathrm{mean}}-\mu_{2}C_{\mathrm{min}}\label{eq:JC1}
\end{equation}
where $\mathbf{i}=(\mathfrak{i}_{1},\ldots,\mathfrak{i}_{N_{O}})$
and $\mathbf{\Omega}=(\Omega_{1},\ldots,\Omega_{N_{O}})$ are the
decision variables, and $\mu_{i}$ are weights chosen depending on
the goals of the constellation design. An optimal coverage is obtained
by solving the optimization problem
\begin{equation}
\begin{alignedat}{1} & (\mathbf{i}^{*},\mathbf{\Omega}^{*})=\arg\min_{\mathbf{i},\mathbf{\Omega}}J_{C}(\mathbf{i},\mathbf{\Omega})\\
 & \mathrm{subject\;to}:\\
 & \mathbf{i}\in D_{\mathfrak{i}}^{N_{O}},\;\mathbf{\Omega}\in D_{\Omega}^{N_{O}}.
\end{alignedat}
\label{eq:optC1}
\end{equation}
Problem (\ref{eq:JC1}) is nonlinear and non-convex, and the considerations
of Remark \ref{rem:comp_compl_0} hold. However, differently from
(\ref{eq:opt0}), minimization is made only w.r.t. $\mathbf{i}$ and
$\mathbf{\Omega}$, and not w.r.t. $\boldsymbol{\theta}_{0}$, which
here is assumed null. Such a dimension reduction of the search domain
allows us to decrease the runtime needed to solve the problem. Based
on an optimal coverage obtained from the solution of (\ref{eq:optC1}),
the constellation revisit times are minimized in Stage 2 by solving
a ``fast'' Linear Programming (LP) problem, as shown next.

\subsection{Stage 2: Revisit time minimization}

\label{subsec:Stage-2}

Starting from the orbits designed in Stage 1, in Stage 2 we reduce
the revisit times by optimizing the constellation w.r.t. the initial
True Anomalies (TAs). 

From Stage 1, we obtain the sets $\bar{\mathcal{T}}_{j}$, $j=1,\ldots,N_{T}$
containing the time instants $t_{j,i}$ at which the targets $j=1,\ldots,N_{T}$
are visited by the satellites of the constellation in a sidereal day
(see (\ref{eq:Tb})).  The times $t_{j,i}\in\bar{\mathcal{T}}_{j}$
have been computed assuming null PAs and null initial TAs: $\omega_{l}=0$,
$\theta_{0l}=0$, $l=1,\ldots,N_{S}$.  The idea is to suitably change
the values of the initial TAs to minimize the constellation revisit
times.

To this aim, suppose that a satellite $l$ starts its orbit with a
given initial TA $\theta_{0l}=-2\pi\tau_{l}/P_{O}$, where $\tau_{l}$
is a time-shift and $P_{O}$ is the orbit period. Let $l_{j,i}$ be
the index of the satellite visiting target $j$ at time $t_{j,i}$
and define the shifted visit time
\[
\breve{t}_{j,i}\doteq t_{j,i}+\tau_{l_{j,i}}
\]
where $t_{j,i}\in\bar{\mathcal{T}}_{j}$, $l_{j,i}\in\{1,\ldots,N_{S}\}$.
Given these time-shifts, we define the following set of revisit times
for target $j$:
\[
\breve{\Delta}_{j}\doteq\{\breve{\delta}{}_{j,1},\breve{\delta}{}_{j,2},\ldots,\breve{\delta}_{j,\mathcal{N}_{j}}\}
\]
where $\breve{\delta}_{j,i}\doteq\breve{t}_{j,i+1}-\breve{t}_{j,i}=t_{j,i+1}-t_{j,i}+\tau_{l_{j,i+1}}-\tau_{l_{j,i}}$,
$i=1,2,\ldots,\mathcal{N}_{j}$. If $\mathcal{N}_{j}=0$ for some
$j$, we set $\breve{\Delta}_{j}\doteq\{\infty\}$. The resulting
maximum revisit times of the constellation in the interval $[0,T_{f}]$
is
\begin{equation}
\breve{\Delta}_{\mathrm{max}}\doteq\max_{j=1:N_{T}}\,\max_{i=1:\mathcal{N}_{j}}\,\breve{\delta}{}_{j,i}.\label{eq:rvt2}
\end{equation}

We now introduce the objective function
\[
J_{\mathrm{rvt}}(\boldsymbol{\tau})\doteq\breve{\Delta}_{\mathrm{max}}
\]
where $\boldsymbol{\tau}=(\tau_{1},\ldots,\tau_{N_{S}})$ and $\mu_{i}$
are weights chosen depending on the goals of the constellation design.
An optimal revisit time configuration is obtained by solving the optimization
problem
\begin{equation}
\begin{alignedat}{1} & \boldsymbol{\tau}^{*}=\arg\min_{\boldsymbol{\tau}}J_{\mathrm{rvt}}(\boldsymbol{\tau})\\
 & \mathrm{subject\;to}:\\
 & \boldsymbol{\tau}\in[\underline{T},\overline{T}]^{N_{S}}
\end{alignedat}
\label{eq:optC2}
\end{equation}
where $\underline{T}$ and $\overline{T}$ are suitable bounds. 

\begin{remark}\label{rem:comp_compl_C2} If the the chronological
ordering of all shifted visits is maintained fixed, under an appropriate
cyclic boundary condition, problem (\ref{eq:optC2}) can be reformulated
as a Linear Programming problem. $\square$ \end{remark}

By solving (\ref{eq:optC2}), we obtain an optimal revisit time configuration.
However, changing the initial TAs implies changing the ground-tracks
of the orbits and consequently loosing the optimal coverage found
in Stage 1. To overcome this issue, we make an adjustment of the RAANs
in such a way that the optimal ground-tracks of Stage 1 are maintained
but the revisit times are reduced according to (\ref{eq:optC2}).
This is accomplished by re-defining the RAANs as
\[
\Omega_{l}:=\Omega_{l}+\tau_{l}\omega_{E},\;l=1,\ldots,N_{S}.
\]

The revisit time optimization approach presented here has been developed
implicitly assuming that each orbit is travelled by only one satellite,
that is, $N_{S}=N_{O}$. However, it may be useful to have multiple
satellites traveling on each orbit, that is, $N_{S}=n_{S}N_{O}$,
where $n_{S}$ is the number of satellites per orbit. In this case,
the same approach can be applied by simply re-defining the sets $\bar{\mathcal{T}}_{j}$
as
\[
\begin{array}{cc}
\bar{\mathcal{T}}_{j}:= & \underbrace{\bar{\mathcal{T}}_{j}\cup\ldots\cup\bar{\mathcal{T}}_{j}}\\
 & n_{S}\;\textrm{times}.
\end{array}
\]

\section{Two-stage constellation optimization with discretized orbital parameters}

\label{sec:2stage_discr}

In this section, we propose an alternative two-stage approach to constellation
optimization in which the admissible domains of the orbital inclination
and right ascension of the ascending node (RAAN) are discretized.
This converts the first-stage continuous, nonconvex optimization problem
into a finite combinatorial problem. The rationale is that a suitably
chosen discretization can facilitate global exploration of the design
space. In particular, a coarse grid does not represent variations
of the objective occurring at scales smaller than the grid spacing
and may therefore exclude narrow attraction basins associated with
poor local minima. Moreover, combinatorial heuristics can move directly
between nonadjacent grid points---or modify several discrete variables
simultaneously---without following a continuous descent trajectory
through the intermediate orbital configurations. These nonlocal moves
can reduce the dependence of the final solution on its initialization
and increase the probability of identifying promising regions of the
design space within a limited computational budget. This principle
is related to coarse-to-fine and graduated-optimization methods, in
which a simplified or coarse-grained problem is solved before progressively
introducing finer details \cite{Hazan2016,Mobahi2015}. It is also
consistent with stochastic combinatorial methods such as simulated
annealing, whose acceptance of non-improving moves enables transitions
between attraction basins \cite{Kirkpatrick1983}. 

Discretization nevertheless introduces a trade-off between computational
complexity and approximation accuracy. A coarse grid produces a smaller
combinatorial problem and promotes broad exploration, but it may fail
to represent the true continuous optimum. Conversely, a fine grid
reduces the discretization error at the cost of a rapidly increasing
number of candidate orbital configurations. Consequently, discretization
should not be interpreted as eliminating nonconvexity or guaranteeing
a globally optimal solution. Rather, it provides a controllable approximation
that makes the problem compatible with efficient combinatorial heuristics
and, when the resulting objective is quadratic in binary decision
variables, with QUBO solvers \cite{Kochenberger2014,QUBO}. A coarse-to-fine
strategy can also be adopted: an initial coarse grid is used to locate
promising regions, which are subsequently refined through a finer
discretization or continuous local optimization.

Stage 1 of the discretized approach is now presented. Stage 2 is exactly
the one of Section \ref{subsec:Stage-2}.

The first operation of the proposed coverage maximization approach
is to discretize the inclination and RAAN domains. For instance, these
domains can be discretized by considering several values distributed
in $D_{\mathfrak{i}}$ and other values distributed in $D_{\Omega}$,
giving rise to a set of points in $D_{\mathfrak{i}}\times D_{\Omega}$.
Another strategy can be to consider classical constellation structures,
like Walker or more general Flower constellations, with different
numbers of orbital planes and different numbers of satellites per
plane. In any case, the umber of orbits can be chosen on the basis
of the desired ground-track resolution. 

From the obtained set of discretized orbital parameters, $N_{O}\in\mathbb{N}$
orbits are generated according to (\ref{eq:ore}) and the corresponding
ground-tracks are computed using (\ref{eq:gt_lonlat}). Then, the
Visit-Count (VC) matrix 
\[
M\doteq[M_{jl}]\in\mathbb{R}^{N_{T}\times N_{O}}
\]
is computed, where the entries $M_{jl}$ are given in (\ref{eq:visit_count}).
We recall that $M_{jl}$ is the number of times that a target $j$
is visited by a satellite $l$ in the time interval of interest.

Now, define the binary vector $\xi=(\xi_{1},\ldots,\xi_{N_{O}})\in\{0,1\}^{N_{O}}$,
where $\xi_{l}=1$ if orbit $l$ is traveled by a satellite, $\xi_{l}=0$
otherwise. The $j$th element of the vector $M\xi\in\mathbb{N}_{0}^{N_{T}}$
gives the total count $\mathcal{N}_{j}$ of visits that target $j$
receives in the time interval of interest. That is, $\mathcal{N}_{j}=(M\xi)_{j}=M_{j:}\xi$.
According to (\ref{eq:cove1}), the mean and minimum constellation
coverages are given by
\[
\begin{alignedat}{1} & C_{\mathrm{mean}}\doteq\frac{1}{N_{T}}\sum_{j=1}^{N_{T}}\mathcal{N}_{j}=\frac{1}{N_{T}}\sum_{j=1}^{N_{T}}M_{j:}\xi\\
 & C_{\mathrm{min}}\doteq\min_{j=1,\ldots,N_{T}}\mathcal{N}_{j}=\min_{j=1,\ldots,N_{T}}M_{j:}\xi.
\end{alignedat}
\]
We can now define the objective function
\begin{equation}
J_{D}(\xi)\doteq-\mu_{1}C_{\mathrm{mean}}-\mu_{2}C_{\mathrm{min}}+\Lambda\left(\sum_{l=1}^{N_{O}}\xi_{l}-N_{\mathrm{des}}\right)^{2}\label{eq:JD}
\end{equation}
where $N_{\mathrm{des}}$ is the desired number of satellites in the
constellation and $\Lambda\geq0$ is sufficiently large. The first
and second terms in (\ref{eq:JD}) aim at maximizing the constellation
coverage. The third term enforces the number of employed orbits to
be close to $N_{\mathrm{des}}$. 

An optimal coverage is obtained by solving the following optimization
problem:
\begin{equation}
\xi^{*}=\arg\min_{\xi\in\{0,1\}^{N_{O}}}J_{D}(\xi).\label{eq:optD1}
\end{equation}
\begin{remark}Problem (\ref{eq:optD1}) is an integer quadratic programming
problem. Finding a global solution involves an exponential/combinatorial
computational complexity. Local solutions can be found using classical
optimization algorithms like genetic algorithms, simulated annealing,
surrogate optimization. $\square$ \end{remark}

\section{QUBO-based constellation design}

In this section, we propose a QUBO-based method for constellation
optimization, following the two-stage approach of Section (\ref{sec:2stage_discr}).
While Stage 2 of this method is exactly the one of Section \ref{subsec:Stage-2},
Stage 1 relies on a QUBO (Quadratic Unconstrained Binary Optimization)
formulation of problem (\ref{eq:optD1}). In this formulation, the
cost function is of the form $\xi^{\top}Q\,\xi$ , where $\xi$ is
a vector of binary variables and $Q$ is a real square matrix. The
QUBO formulation is of interest since it enables the use of efficient
QUBO solvers and quantum annealers, a particular class of quantum
computers that are nowadays commercially available \cite{kadowaki1998quantum,Johnson2011,dwave_site}.
Even more importantly, it leads to a significant simplification of
the optimization problem, where closed-form solutions can be found.

\subsection{QUBO formulation}

\label{subsec:QUBO-formulation}

We show that the three terms of the cost function $J_{D}$ in (\ref{eq:JD})
can be recast into the QUBO form. 

The first is proportional to $C_{\mathrm{mean}}$, which can be written
as

\[
\begin{alignedat}{1}C_{\mathrm{mean}} & =\frac{1}{N_{T}}\sum_{j=1}^{N_{T}}M_{j:}\xi=\frac{1}{N_{T}}\mathbf{1}^{\top}M\xi\\
 & =\frac{1}{N_{T}}\mathbf{1}^{\top}M\xi^{2}=\xi^{\top}W_{M}\xi
\end{alignedat}
\]
where 
\[
W_{M}\doteq\frac{1}{N_{T}}\mathrm{diag}(\mathbf{1}^{\top}M)
\]
$\mathbf{1}\equiv\mathbf{1}_{N_{T}\times1}$ and $\xi^{2}$ is the
entry-wise square of $\xi$. Note that $\xi^{2}=\xi$ as the entries
of $\xi$ are binary.

The second term of $J_{D}$ is $-\mu_{2}C_{\mathrm{min}}$ and it
is aimed at increasing the minimum coverage. This term cannot be directly
written in the QUBO form as the $\mathrm{min}$ operation is not a
quadratic function. This operation could be treated by introducing
linear constraints and additional auxiliary variables, leading to
an increased-complexity procedure and optimization problem. To overcome
these issues, we propose two options, where quadratic functions are
introduced, aimed at promoting orbits that visit targets with small
visit counts compared to others.

\emph{Option 1. Weighted linear coverage surrogate}. Let 
\[
\overline{C}_{j}=\sum_{l=1}^{N_{O}}M_{jl}
\]
denote the maximum number of visits potentially available to target
$j$, and define 
\[
\eta_{j}=\frac{1}{\left(\varepsilon+\overline{C}_{j}\right)^{p}},
\]
where $\varepsilon>0$ prevents division by zero and $p>0$ controls
the weighting intensity. A QUBO-compatible surrogate for the minimum
coverage is 
\[
\hat{C}_{\min}(\xi)=\sum_{j=1}^{N_{T}}\eta_{j}M_{j:}\xi=\eta^{\top}M\xi.
\]
Since the entries of $\xi$ are binary, this expression can be written
as 
\[
\hat{C}_{\min}(\xi)=\xi^{\top}W_{S}\,\xi,\qquad W_{S}\doteq\operatorname{diag}\!\left(\eta^{\top}M\right).
\]
This surrogate favors orbits covering targets with fewer available
observation opportunities. Moreover, $W_{S}$ is diagonal, so the
formulation requires no auxiliary variables and introduces no additional
couplings in the QUBO problem.

\emph{Option 2. Mean--variance surrogate}. Let $\mathbf{1}\equiv\mathbf{1}_{N_{T}\times1}$
and
\[
P=I_{N_{T}}-\frac{1}{N_{T}}\mathbf{1}\mathbf{1}^{\top}.
\]
The coverage dispersion is 
\[
V_{C}(\xi)=\frac{1}{N_{T}}\|PM\xi\|^{2}=\xi^{\top}W_{V}\xi,\qquad W_{V}=\frac{1}{N_{T}}M^{\top}PM.
\]
A QUBO-compatible surrogate for the minimum-coverage contribution
is then defined as 
\[
\hat{C}_{\min}(\xi)=-V_{C}(\xi)=\xi^{\top}W_{S}\xi,\qquad W_{S}\doteq-W_{V}.
\]
Maximizing $\hat{C}_{\min}$ favors constellations with more uniformly
distributed target coverage and thus discourages poorly covered targets.
The surrogate requires no auxiliary binary variables, although $W_{S}$
may be dense.

The third term of $J_{D}$ is proportional to $\left(\sum_{l=1}^{N_{O}}\xi_{l}-N_{\mathrm{des}}\right)^{2}$,
which can be written as
\[
\begin{array}{c}
\left(\sum_{l=1}^{N_{O}}\xi_{l}-N_{\mathrm{des}}\right)^{2}=\left(\mathbf{1}_{1\times N_{O}}\xi-N_{\mathrm{des}}\right)^{2}\vs{2mm}\\
=\xi^{\top}\mathbf{1}_{N_{O}}\xi-2N_{\mathrm{des}}\mathbf{1}_{1\times N_{O}}\xi+N_{\mathrm{des}}^{2}\vs{2mm}\\
=\xi^{\top}W_{O}\xi+N_{\mathrm{des}}^{2}
\end{array}
\]
where
\[
W_{O}\doteq\mathbf{1}_{N_{O}}-2N_{\mathrm{des}}I_{N_{O}}.
\]
The matrix $W_{O}$ allows us to impose a prescribed number $N_{\mathrm{des}}$
of satellites in the constellation. 

Now, we define the objective function 
\[
\begin{alignedat}{1} & J_{Q}(\xi)\doteq\xi^{\top}Q\:\xi\\
 & Q\doteq-\mu_{1}W_{M}-\mu_{2}W_{S}+\Lambda W_{O}.
\end{alignedat}
\]
An optimal coverage is obtained by solving the following optimization
problem:
\begin{equation}
\xi^{*}=\arg\min_{\xi\in\{0,1\}^{N_{O}}}J_{Q}(\xi).\label{eq:optQ1}
\end{equation}

\begin{remark}Problem (\ref{eq:optQ1}) is an integer quadratic programming
problem. Finding a global solution involves an exponential/combinatorial
computational complexity. Satisfactory local solutions can be found
using efficient QUBO solvers. $\square$ \end{remark}

\begin{remark}Under certain conditions, quantum annealers may be
able to find global solutions of (\ref{eq:optQ1}). On the other hand,
quantum annealers my struggle to solve the problem since $W_{O}$
and $W_{S}$ (Option 2) are dense matrices. This issue may be overcome
by using Option 1 and replacing $\Lambda W_{O}$ with the simple term
$\Lambda\sum_{l=1}^{N_{O}}\xi_{l}$. This terms allows us to minimize
the number of satellites but implies a more difficult tuning of $\Lambda$.
$\square$ \end{remark}

\subsection{QUBO closed-form solution}

\label{sec:qubo_closed_form}

When the weighted linear coverage surrogate introduced in Section
\ref{subsec:QUBO-formulation} is adopted, both $W_{M}$ and $W_{S}$
are diagonal. This structure allows the prescribed number of satellites
to be enforced directly through a selection rule, without introducing
the matrix $W_{O}$ or tuning the penalty coefficient $\Lambda$.

Let $N_{S}=N_{\mathrm{des}}$, with $0\leq N_{S}\leq N_{O}$, and
define 
\[
Q_{C}=-\mu_{1}W_{M}-\mu_{2}W_{S}=\operatorname{diag}(q_{1},\ldots,q_{N_{O}}).
\]
The coverage optimization problem with exactly $N_{S}$ satellites
can then be written as 
\begin{equation}
\min_{\xi\in\{0,1\}^{N_{O}}}\xi^{\top}Q_{C}\xi\qquad\text{subject to}\qquad\sum_{l=1}^{N_{O}}\xi_{l}=N_{S}.\label{eq:coverage_fixed_cardinality}
\end{equation}
Since $\xi_{l}^{2}=\xi_{l}$, the objective reduces to 
\[
\xi^{\top}Q_{C}\xi=\sum_{l=1}^{N_{O}}q_{l}\xi_{l}.
\]
In particular, the diagonal coefficients are 
\[
q_{l}=-\frac{\mu_{1}}{N_{T}}\sum_{j=1}^{N_{T}}M_{jl}-\mu_{2}\sum_{j=1}^{N_{T}}\eta_{j}M_{jl},\qquad l=1,\ldots,N_{O}.
\]
Thus, each candidate orbit has a fixed contribution to the objective,
independently of the other selected orbits. Let $\pi$ be a permutation
of the orbit indices such that 
\[
q_{\pi(1)}\leq q_{\pi(2)}\leq\cdots\leq q_{\pi(N_{O})}.
\]
A global solution of~(\ref{eq:coverage_fixed_cardinality}) is given
by 
\begin{equation}
\xi_{\pi(k)}^{\star}=\begin{cases}
1, & k=1,\ldots,N_{S},\\
0, & k=N_{S}+1,\ldots,N_{O}.
\end{cases}\label{eq:coverage_selection}
\end{equation}
Indeed, replacing a selected orbit with an unselected orbit having
a smaller coefficient cannot increase the objective. Consequently,
selecting the $N_{S}$ smallest diagonal entries yields the global
minimum. Ties can be resolved arbitrarily.

\begin{remark}The resulting procedure requires only the computation
of the diagonal coefficients and their sorting, whose computational
complexity is $O(N_{O}\log N_{O})$. It therefore avoids the dense
cardinality-penalty matrix, auxiliary variables, and numerical QUBO
optimization, while enforcing the satellite count exactly. $\square$
\end{remark}

\begin{remark}The optimality guarantee concerns the fixed-cardinality
problem with the weighted linear coverage surrogate. Since this surrogate
approximates the minimum-coverage criterion, the selected constellation
is not necessarily optimal for the original minimum coverage. Furthermore,
the selection rule does not generally apply to the mean--variance
surrogate, whose off-diagonal terms introduce interactions between
orbit selections. $\square$ \end{remark}

\section{Simulation results}

Numerical experiments have been carried out for testing the two-stage
approach and comparing it with the general nonlinear approach and
a standard design based on classical Wlaker Delta constellations.
The results presented here are preliminary and regard the case of
continuous orbital parameters; more comprehensive simulation campaigns
are currently in progress regarding also the case of discretized orbital
parameters. 

A set of $N_{T}\in[10,100]$ targets was considered, randomly distributed
on the Earth surface, with latitude in $[-60^{o},60^{o}]$ and longitude
in $[-180^{o},180^{o}]$. Constellations composed of $N_{S}\in[6,25]$
satellites were adopted, with Repeat Ground Track (RGT) orbits of
altitude $554.38\;\mathrm{km}$ ($15$ revolutions per day). The constellations
were designed using the general nonlinear approach (directly solving
(\ref{eq:opt0})), the two-stage approach of Section \ref{sec:2st_optim}
and a classical Walker approach, considering different numbers of
orbital planes and satellites per plane. Numerical tests using the
discretized and QUBO approaches are underway. The goal of the design
was to minimize the maximum revisit time (Max RVT), while ensuring
a $100\%$ coverage of the targets. 

All the optimization problems were solved using the Particle Swarm
algorithm of the Matlab Global Optimization Toolbox, running on a
workstation with the following features: CPU Intel Xeon w7-2575X,
base speed 3 GHz, max speed 4.8 GHz, 22 cores, 44 logical processors;
GPU NVIDIA RTX PRO 4500 Blackwell, 32 GB dedicated memory; 256 GB
RAM; hard disk 4 TB. 

\begin{figure}[h]
\begin{centering}
\includegraphics[scale=0.35]{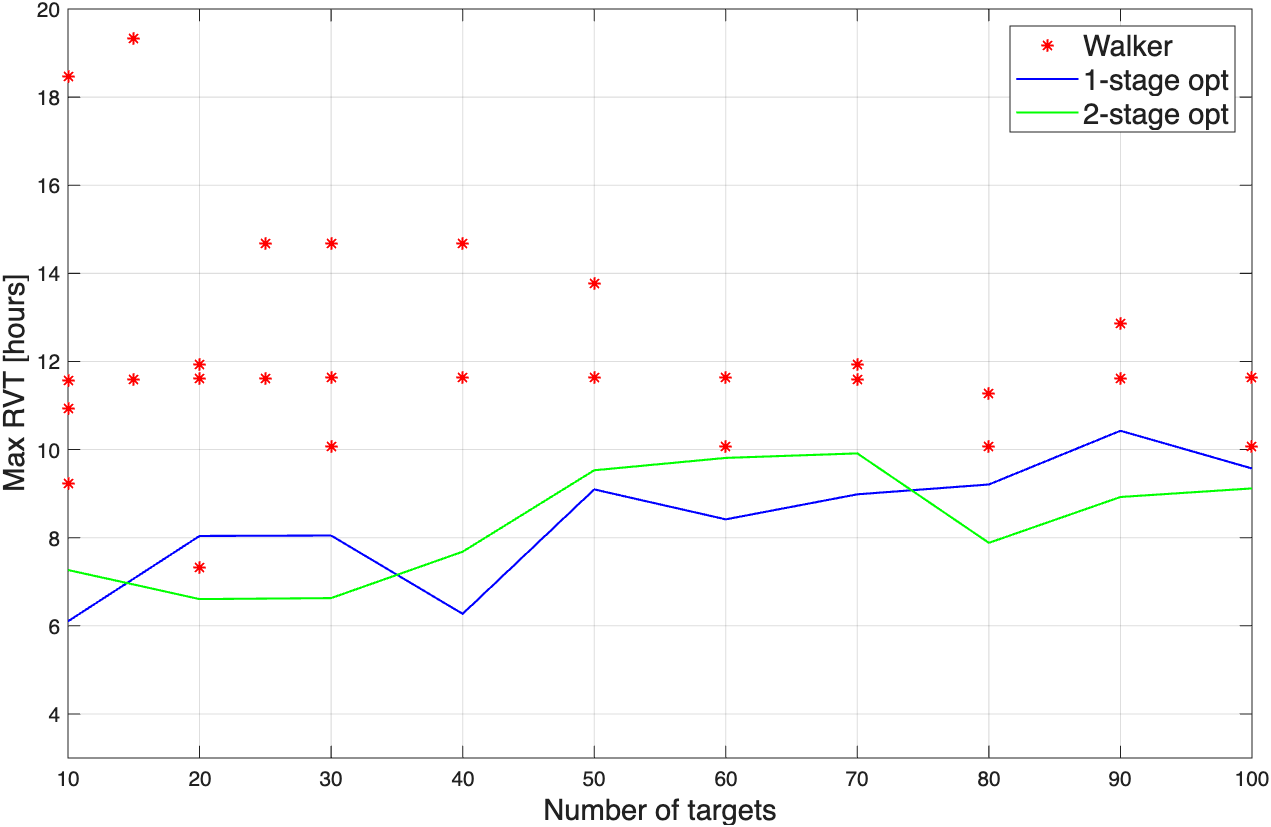}
\par\end{centering}
\caption{Maximum revisit times $[\mathrm{hours}]$ for $N_{S}=10$ and $N_{T}\in[10,100]$.
\label{fig:rt_1}}

\end{figure}

Table \ref{tab:rt_1} provides the mean and maximum runtimes required
by the 1-stage and 2-stage approaches (called 1STG and 2STG) to solve
the minimization problem yielding the optimal constellation design.
Mean and maximum are computed over the set of $N_{T}\in[10,100]$
targets.\textcolor{red}{{} }Figure \ref{fig:rt_1} shows the Max RVT
in function of the number of targets, using Walker constellations,
and constellations designed using the 1-stage and 2-stage approaches.
In the figure, the results obtained with $N_{S}=10$ satellites are
shown. Note that these results strongly depend on the angle of aperture
assumed for the sensor mounted on the satellites. In these simulations,
an aperture half-angle of $37^{o}$ was adopted, corresponding to
a circular region on the Earth surface with radius $\sim433\;\mathrm{km}$.
$ $

\begin{table}[h]
\begin{centering}
\begin{tabular}{ccccc}
NS & \shortstack{1STG\\mean} & \shortstack{2STG\\mean} & \shortstack{1STG\\max} & \shortstack{2STG\\max} \\ 
\hline 
10 & 99.6 & 92.8 & 165.1 & 145.9 \\ 
15 & 191.2 & 181.2 & 389.7 & 288.7 \\ 
25 & 419.6 & 287.3 & 677 & 615.9 \\ 
\hline 
\end{tabular}
\par\end{centering}
\caption{Optimization runtimes $[\mathrm{s}]$ of 1-stage and 2-stage approaches.
\label{tab:rt_1}}
\end{table}

\bibliographystyle{IEEEtran}
\addcontentsline{toc}{section}{\refname}\bibliography{large_constellations,quantum_comp,optimization}

@article{Stoilova2022,
	author = {Krasimira Stoilova and Todor Stoilov},
	doi = {10.3390/app12094147},
	issn = {20763417},
	issue = {9},
	journal = {Applied Sciences (Switzerland)},
	title = {Model Predictive Traffic Control by Bi-Level Optimization},
	volume = {12},
	year = {2022}}

@article{Lagona2022,
	author = {Enrico Lagona and Samuel Hilton and Andoh Afful and Alessandro Gardi and Roberto Sabatini},
	doi = {10.1016/j.actaastro.2022.01.027},
	issn = {00945765},
	journal = {Acta Astronautica},
	title = {Autonomous Trajectory Optimisation for Intelligent Satellite Systems and Space Traffic Management},
	volume = {194},
	year = {2022}}

@article{const_ps_2024,
	author = {Y. Yang and X. Wu and J. Li and J. Zang and J. Lu and R. Zgeib},
	journal = {Space: Science {\&} Technology},
	title = {Configuration Design Method of Mega Constellation for Low Earth Orbit Observation},
	volume = {4},
	year = {2024}}

@article{Appel2014,
	author = {Leonid Appel and Moshe Guelman and David Mishne},
	doi = {10.1016/j.actaastro.2014.02.016},
	issn = {00945765},
	issue = {1},
	journal = {Acta Astronautica},
	title = {Optimization of satellite constellation reconfiguration maneuvers},
	volume = {99},
	year = {2014}}

@misc{Curzi2020,
	author = {Giacomo Curzi and Dario Modenini and Paolo Tortora},
	doi = {10.3390/AEROSPACE7090133},
	issn = {22264310},
	issue = {9},
	journal = {Aerospace},
	title = {Large constellations of small satellites: A survey of near future challenges and missions},
	volume = {7},
	year = {2020}}

@article{Thangavel2024,
	author = {Kathiravan Thangavel and Raja Pandi Perumal and Khaja Faisal Hussain and Alessandro Gardi and Roberto Sabatini},
	doi = {10.1016/j.actaastro.2023.12.055},
	issn = {00945765},
	journal = {Acta Astronautica},
	title = {Multidisciplinary design and optimization of intelligent Distributed Satellite Systems for earth observation},
	volume = {216},
	year = {2024}}

@misc{Zhang2022,
	author = {Zhang Zhang and Lei Deng and Jiaqi Feng and Liang Chang and Dong Li and Yilin Qin},
	doi = {10.3390/aerospace9070362},
	issn = {22264310},
	issue = {7},
	journal = {Aerospace},
	title = {A Survey of Precision Formation Relative State Measurement Technology for Distributed Spacecraft},
	volume = {9},
	year = {2022}}

@inproceedings{Nunes2013,
	author = {Miguel A. Nunes},
	booktitle = {Advances in the Astronautical Sciences},
	issn = {00653438},
	title = {Satellite constellation optimization method for future earth observation missions using small satellites},
	volume = {146},
	year = {2013}}

@article{Crisp2019,
	author = {N. H. Crisp and K. L. Smith and P. M. Hollingsworth},
	doi = {10.1017/aer.2019.57},
	issn = {00019240},
	issue = {1266},
	journal = {Aeronautical Journal},
	title = {An integrated design methodology for the deployment of constellations of small satellites},
	volume = {123},
	year = {2019}}

@article{Melaku2023,
	author = {Shimeles Demissie Melaku and Hae Dong Kim},
	doi = {10.3390/rs15061572},
	issn = {20724292},
	issue = {6},
	journal = {Remote Sensing},
	title = {Optimization of Multi-Mission CubeSat Constellations with a Multi-Objective Genetic Algorithm},
	volume = {15},
	year = {2023}}

@article{Deccia2022,
	author = {Carlos M.A. Deccia and David N. Wiese and Robert S. Nerem},
	doi = {10.3390/rs14143340},
	issn = {20724292},
	issue = {14},
	journal = {Remote Sensing},
	title = {Using a Multiobjective Genetic Algorithm to Design Satellite Constellations for Recovering Earth System Mass Change},
	volume = {14},
	year = {2022}}

@article{Imoto2023,
	author = {Yuta Imoto and Satoshi Satoh and Toshihiro Obata and Katsuhiko Yamada},
	doi = {10.1016/j.actaastro.2023.02.040},
	issn = {00945765},
	journal = {Acta Astronautica},
	title = {Optimal constellation design based on satellite ground tracks for Earth observation missions},
	volume = {207},
	year = {2023}}

@article{Kim2019,
	author = {Hae Dong Kim and Jae Dong Seong},
	doi = {10.1007/s42405-018-0129-9},
	issn = {20932480},
	issue = {2},
	journal = {International Journal of Aeronautical and Space Sciences},
	title = {Multi-objective Heuristic Design Approach for SAR Mission for Monitoring Local Target Area},
	volume = {20},
	year = {2019}}

@article{Chen2024,
	author = {Junyu Chen and Baolin Wu and Zhaobo Sun and Danwei Wang},
	doi = {10.1016/j.actaastro.2023.10.012},
	issn = {00945765},
	journal = {Acta Astronautica},
	title = {Distributed safe trajectory optimization for large-scale spacecraft formation reconfiguration},
	volume = {214},
	year = {2024}}

@article{Jiaxin2021,
	author = {Hu Jiaxin and Yang Leping and Huang Huan and Zhu Yanwei},
	doi = {10.23919/JSEE.2021.000128},
	issn = {16711793},
	issue = {6},
	journal = {Journal of Systems Engineering and Electronics},
	title = {Optimal reconfiguration of constellation using adaptive innovation driven multiobjective evolutionary algorithm},
	volume = {32},
	year = {2021}}

@article{Arnas2022,
	author = {David Arnas and Richard Linares},
	doi = {10.2514/1.G006514},
	issn = {15333884},
	issue = {7},
	journal = {Journal of Guidance, Control, and Dynamics},
	title = {Uniform Satellite Constellation Reconfiguration},
	volume = {45},
	year = {2022}}

@article{Fakoor2016,
	author = {Mahdi Fakoor and Majid Bakhtiari and Mahshid Soleymani},
	doi = {10.1016/j.asr.2016.04.031},
	issn = {18791948},
	issue = {3},
	journal = {Advances in Space Research},
	title = {Optimal design of the satellite constellation arrangement reconfiguration process},
	volume = {58},
	year = {2016}}

@article{Soleymani2019,
	author = {Mahshid Soleymani and Mahdi Fakoor and Majid Bakhtiari},
	doi = {10.1016/j.asr.2019.02.003},
	issn = {18791948},
	issue = {10},
	journal = {Advances in Space Research},
	title = {Optimal mission planning of the reconfiguration process of satellite constellations through orbital maneuvers: A novel technical framework},
	volume = {63},
	year = {2019}}

@inproceedings{Hazan2016,
	author = {Hazan, Elad and Levy, Kfir Y. and Shalev-Shwartz, Shai},
	booktitle = {Proceedings of the 33rd International Conference on Machine Learning},
	pages = {1833--1841},
	series = {Proceedings of Machine Learning Research},
	title = {On Graduated Optimization for Stochastic Non-Convex Problems},
	volume = {48},
	year = {2016}}

@inproceedings{Mobahi2015,
	author = {Mobahi, Hossein and Fisher, John W.},
	booktitle = {Energy Minimization Methods in Computer Vision and Pattern Recognition},
	pages = {71--84},
	publisher = {Springer},
	title = {Coarse-to-Fine Minimization of Some Common Nonconvexities},
	year = {2015}}

@article{Kochenberger2014,
	author = {Kochenberger, Gary and Hao, Jin-Kao and Glover, Fred and Lewis, Mark and L{\"u}, Zhipeng and Wang, Haibo and Wang, Yang},
	doi = {10.1007/s10878-014-9734-0},
	journal = {Journal of Combinatorial Optimization},
	pages = {58--81},
	title = {The Unconstrained Binary Quadratic Programming Problem: A Survey},
	volume = {28},
	year = {2014}}

@article{QUBO,
	author = {Fred Glover and Gary Kochenberger and Yu Du},
	journal = {4OR},
	month = {12},
	number = {4},
	pages = {335-371},
	title = {{Quantum Bridge Analytics I: a tutorial on formulating and using QUBO models}},
	volume = {17},
	year = 2019}

@misc{dwave_site,
	title = {{D-Wave Systems}, www.dwavesys.com}}

@article{Kirkpatrick1983,
	author = {S. Kirkpatrick and C. D. Gelatt and M. P. Vecchi},
	journal = {Science},
	number = {4598},
	pages = {671-680},
	title = {Optimization by Simulated Annealing},
	volume = {220},
	year = {1983}}

@article{kadowaki1998quantum,
	author = {Kadowaki, Tadashi and Nishimori, Hidetoshi},
	journal = {Physical Review E},
	number = {5},
	pages = {5355},
	publisher = {APS},
	title = {Quantum annealing in the transverse Ising model},
	volume = {58},
	year = {1998}}

@article{Johnson2011,
	author = {Johnson, Mark W and Amin, Mohammad HS and Gildert, Suzanne and Lanting, Trevor and Hamze, Firas and Dickson, Neil and Harris, Richard and Berkley, Andrew J and Johansson, Jan and Bunyk, Paul and others},
	journal = {Nature},
	number = {7346},
	pages = {194--198},
	publisher = {Nature Publishing Group UK London},
	title = {Quantum annealing with manufactured spins},
	volume = {473},
	year = {2011}}

\end{document}